\documentclass[twocolumn,12pt]{aastex62}

\usepackage{CJK} 
\usepackage{bm}
\usepackage{amsmath}

\begin{document}

\title{Rotation of Two Micron All Sky Survey  Clumps in Molecular Clouds}

\correspondingauthor{Xuefang Xu, Di Li,Y.Sophia Dai}
\email{xfxu@nao.cas.cn}
\email{dili@nao.cas.cn}
\email{ydai@nao.cas.cn}

\author{Xuefang Xu
\begin{CJK}{UTF8}{bsmi}
(徐雪芳)
\end{CJK}}
\affil{CAS Key Laboratory of FAST, National Astronomical Observatories, Chinese Academy of Sciences, Beijing 100101, China} 
\affil{University of Chinese Academy of Sciences, Beijing 100049, China}

\author{Di Li
\begin{CJK}{UTF8}{bsmi}
(李菂)
\end{CJK}}
\affil{CAS Key Laboratory of FAST, National Astronomical Observatories, Chinese Academy of Sciences, Beijing 100101, China} 
\affil{University of Chinese Academy of Sciences, Beijing 100049, China}
\affil{NAOC-UKZN Computational Astrophysics Centre, University of KwaZulu-Natal, Durban 4000, South Africa}

\author{Y.Sophia Dai 
\begin{CJK}{UTF8}{bsmi}
(戴昱)
\end{CJK}}
\affil{Chinese Academy of Sciences South America Center for Astronomy (CASSACA), 
NAOC, Beijing 100101, China}

\author{Paul F.~Goldsmith}
\affil{Jet Propulsion Laboratory, California Institute of Technology, Pasadena, CA 91109, USA}

\author{Gary A.~Fuller}
\affiliation{Jodrell Bank Centre for Astrophysics, Department of Physics and Astronomy, the University of Manchester, Oxford Road, Manchester, M13 9PL, UK}
              
\begin{abstract}
We have analyzed the rotational properties of 12 clumps using $^{13}$CO (1--0) and 
C$^{18}$O (1--0) maps of the Five College Radio Astronomy Observatory 
13.7 m radio telescope. The clumps, located within molecular clouds, 
have radii ($R$) in the range of 0.06 -- 0.27\,pc. 
The direction of clump elongation is not correlated with the direction of the velocity gradient. 
We measured the specific angular momentum (J/M) to be between 0.0022 and 
0.025 pc\,km\,s$^{-1}$ based on $^{13}$CO images, 
and between 0.0025 and 0.021 pc\,km\,s$^{-1}$ based on C$^{18}$O images. 
The consistency of $J/M$ based on different tracers indicates the $^{13}$CO 
and C$^{18}$O in dense clumps trace essentially the same material 
despite significantly different opacities.
We also found that $J/M$ increases monotonically as a function of $R$ in power--law form, 
$J/M~\propto~R^{1.58~\pm~0.11}$. 
The ratio between rotation energy and gravitational energy, $\beta$, 
ranges from 0.0012 to 0.018. 
The small values of $\beta$ imply that rotation alone is not sufficient 
to support the clump against gravitational collapse.
\end{abstract}

             
\section{Introduction} \label{sec:intro}
Molecular clouds are the major active star--forming regions 
in the Milk Way~\citep[]{zuckerman1974,burton1976}. 
Within are denser condensations (referred to as clumps) of size $\simeq$ 0.1 pc, 
which are are often gathered into larger-scale 
filamentary structures~\citep[see e.g.][]{francesco2007}. 
It is still arguable whether stars are born in a single system 
or in multiple systems~\citep[see e.g.][]{larson2010}. 
To understand the star-formation process, 
we really must trace the evolution of gas and dust starting at the scale of clouds, 
through clumps, down to the smaller scale of cores and finally to stars and stellar clusters. 
Molecular line emission plays a crucial role in the star formation process, 
as it is the primary coolant of gas at densities $\leq$ $\simeq$ 10$^6$ cm$^{-3}$ 
when dust--gas coupling starts to dominate 
the thermal behavior~\citep[e.g.][]{goldsmith2001,sipila2012}. 
Wide-field maps of molecular lines offer rich kinematic information, 
based on the line profiles and their variation, 
but the main challenge for kinematic studies is the variation of molecular opacity, 
excitation, and abundance along the line of sight. 

Thus, different molecular lines are chosen for different science goals. 
For example, the velocity gradient of N$_{2}$H$^{+}$ (1-0) emission 
was used to investigate the angular momentum of 
cores in the Orion A cloud~\citep{tatematsu2016}.
NH$_{3}$ inversion transitions (1,1) and (2,2) were used to derive the 
kinetic temperature of cores in the OMC2 and OMC3 regions of Orion~\citep{di2013}.
A high angular and velocity resolution C$^{18}$O (1-0) emission map 
was used to get an initial overview of the filamentary structure 
in the Orion A molecular cloud~\citep{suri2019}. 

Rotation is one of the most fundamental physical parameters of cloud clumps. 
Besides affecting the mass distribution of the cloud clumps, 
rotation has the following important effects.
First, it could provide additional support against collapse. 
For example, the rotation was left as a free parameter to fit emission spectra 
in~\citet{adams1987} models of  young stellar objects (YSOs). 
Second, rotation can also twist magnetic-field lines by transferring 
the angular momentum from the cloud to the surrounding medium. 
This interaction between rotation and magnetic field was used to explain the 
angular momentum problem~\citep[e.g.,][]{field1978,mouschovias1979,mouschovias1987}: 
the angular momentum of prestellar cores is many orders of magnitude 
larger than that can be contained within a single star, 
even though cores are observed to be rotating much less than originally predicted. 
Finally, when coupled with cloud morphology, 
the rotation rate affects possible  fragmentation 
(i.e., formation of single or multiple prestellar objects)
~\citep[e.g.,][]{bodenheimer1980, boss1993, xiao2018}. 

 $\beta$ is the ratio between rotational energy and gravitational energy, 
 which can be used to infer the dynamical status of cores, clumps and molecular clouds. 
In general, a large $\beta$ ($\geqslant$ 0.25--0.3) implies a stable gas cloud, 
which is not fragmenting and thus no longer forming stars.  
In a numerical simulation,~\citet{ostriker1973} found that  with $\beta{}={}0.25$ 
gravity is initially nearly balanced by rotation.
On the other hand, a small $\beta$ indicates instabilities against gravitational collapse.
So far, this is the case for observed $\beta$ values
~\citep[see e.g.][]{goodman1993,caselli2002,chen2009,li2012,tatematsu2016}. 
Several theoretical models have predicted specific $\beta$ values 
to be associated with specific events, 
such as the formation of bars or rings ($\beta$ = 0.01)
~\citep[see e.g.][]{bodenheimer1978,
rozyczka1980,stahler1983}.
~\citet{boss1999} reported that rotating, magnetized cloud cores 
initially fragment when $\beta${}\textless{}0.1.

One way to measure the rotation is through large-scale velocity gradients. 
Large-scale velocity gradients are detected from molecular line emission maps. 
These gradients can be explained by rotation, outflow, infall, 
or motions between smaller unresolved clumps. 
Although it is difficult to distinguish definitively between these scenarios,
the majority of previous researches have attributed observed 
velocity gradients to rotation~\citep[e.g.][]{clark1982,kane1997,pirogov2003,redman2004,
shinnaga2004,chen2007,chen2009,li2012}. 
In our work, line wings were not detected and in consequence, 
outflows were not further considered.
In general, velocity gradients can be obtained by linear fitting of velocity field
($ v_{LSR} = v_0 + c_{1}{}\Delta{}\alpha + c_{2}{}\Delta{}\delta $)
~\citep{goodman1993}, 
where the complex nature of the velocity field is ignored. 
It is assumed that clumps follow rigid-body rotation law 
and that the angular velocity is approximately the gradient of the line of sight velocity.  
This is because turbulence is dominant in a larger scale than 
the size of clumps~\citep[e.g.][]{klapp2014}.
Thus rigid-body rotation is an appropriate approximation of the dense clumps. 

Dominantly prograde velocity gradients are observed 
in the~\citet{colombo2014} cloud sample. 
Similar results were found by~\citet{braine2018} for M 33,
and~\citet{braine2020} for M 51.
The  orientation of rotation axis in some studies has been found to be random
~\citep[see e.g.][]{goodman1993,caselli2002,curtis2011,li2012,tatematsu2016}, 
or not, depending on the origin of the initial angular momentum of clouds.  
There are three or four orders of magnitude between the scale of the galaxies 
and that of dense clumps. 
Any robust relation was yet to be established between 
the rotation of dense clumps and the rotation of the Milky Way.

To measure the rotation, various tracers are utilized. 
Different molecular lines probes different excitation
in different regions within the molecular clouds. 
In general, CO traces clouds. $^{13}$CO and C$^{18}$O trace clumps.
NH$_{3}$ and N$_{2}$H$^{+}$ trace cores. 
For instance, utilizing the $^{13}$CO(1--0) map,~\citet{arquilla1986} found that molecular clouds B163 and B163SW rotate sufficiently relatively rapidly that the rotation is affecting their structure. 
Using C$^{18}$O(3--2) map,~\citet{curtis2011} studied clumps in the Perseus molecular cloud and found a dynamically insignificant rotation. 
~\citet{caselli2002} reached the same conclusion by mapping cores with N$_{2}$H$^{+}$(1--0). 
Observationally, a universal power-law relation is found in dense cores/clumps
between the specific angular momentum ($J/M$ = angular momentum/core mass), 
and radius (R). 
$J/M$ increases with R (i.e. $J/M$~$\propto$~R$^{1.6}$), 
for a typical R~$\geqslant$~0.02 pc~\citep{goodman1993}. 
This relation can be used to indicate the initial conditions of the 
cores/clumps in a molecular cloud. 
The power-law relation is similar to Larson's law~\citep{larson1981}.

In this paper, we study the rotation in 12 clumps with the $^{13}$CO(1--0) 
and C$^{18}$O(1--0) maps. 
The clumps were identified using Two Micron All Sky Survey (2MASS) maps. 
The identified clumps are in 6 molecular clouds. 
This paper is organized as follows. 
Section~\ref{sec:data} briefly describes data. 
Section~\ref{sec:sour} introduces the 6 molecular clouds in detail. 
In section~\ref{sec:analyze}, clump identification and velocity gradient fitting are introduced. 
We analyze the specific angular momentum in section~\ref{sec:rotation}. 
Section~\ref{sec:discuss} is the discussion. 
We summarize the main results in section~\ref{sec:sum}.

\section{Data} \label{sec:data}

To study the clumps in molecular clouds,
we observed 6 northern molecular clouds with the Five College Radio Astronomy 
Observatory (FCRAO) 13.7 m radio telescope. 
There are no protostars or YSOs in our clouds. 
The details of the clouds are given in section~\ref{sec:sour}. 
The position and distance of the clouds were summarized in table~\ref{tab:sources}. 
The $J$= 1~$\rightarrow$~0 rotational transitions 
of $^{13}$CO (110.201\,GHz) and C$^{18}$O (109.782\,GHz) 
were observed during 2001 May, 2003 April and May, and 2004 January.

The $^{13}$CO and C$^{18}$O lines were observed simultaneously 
with the on-the-fly (OTF) mapping technique. 
The 15 pixel QUARRY (Quabbin Array Receiver)~\citep{erickson1992} 
was employed to map L1523 and L1257. 
The autocorrelation spectrometer had a bandwidth of 20\,MHz over 1024 channels, 
corresponding to a spectral resolution of 0.065\,km\,/\,s at 110\,GHz. 
The rest of the clouds were observed using the 32 pixel Sequoia receiver. 
The autocorrelation spectrometer had a bandwidth of 25 MHz over 1024 channels, 
corresponding to a spectral resolution of 0.067 \,km\,/\,s at 110\,GHz.  
The angular size of the maps ranged from 15$^{'}$ to 34$^{'}$.
The system temperature was typically 200\,K and the integration time per beam on the sky was 10\,s. 
The full width to half maximum beam widths were  45$^{''}$ and 48$^{''}$ for
$^{13}$CO and C$^{18}$O, respectively.  
To convert intensities from antenna temperature, $T$${_A^{\ast}}$, 
to main beam temperature, $T$$_{mb}$, 
we used $T_{mb} = T{_A^{\ast}}/\eta_{mb}$ with $\eta_{mb} $ = 0.49.

All observations used in this study were carried in a single program. 
The $^{13}$CO and C$^{18}$O maps of L1523 and L1257 
are published for the first time. 
For L1544, B227, L1574, and CB45, 
$^{13}$CO and C$^{18}$O data were shown in~\citet{goldsmith2005}. 
~\citet{zuo2018} used the $^{13}$CO maps of B227, L1574, and CB45 to study  H$_{2}$ formation.

 The 2MASS extinction maps~\citep{dobashi2011} and $^{13}$CO 
 and C$^{18}$O maps have comparable angular resolution. 
 The 2MASS maps were used to identify clumps as discussed in Section~\ref{sec:identi}. 
 Clump masses and sizes were calculated based on the results (the total intensity, 
 semi-major and semi-minor axes of identified clumps) of the identification 
 (see equations (\ref{ident1}) and (\ref{ident2})). 

 \begin{table}[!htbp]
   \caption{Source list. \label{tab:sources}}
 \vspace*{0.5ex}
   \begin{tabular}{l c c c c}
   \hline
   \hline
     {} & {} & {} & D & \\
     Sources & R.A. (J2000) & Dec. (J2000) & (pc) & References \\
     \hline
L1544  & 05:04:18.10 & +25:11:07.6 & 140 & 1 \\
L1523  & 05:06:13.79 & +31:43:59.8 & 140 & 2 \\
B227  & 06:07:28.35 & +19:28:03.8 & 400 & 3 \\
L1574 & 06:08:04.95 & +18:28:12.1 & 300 & 4 \\
CB45 & 06:08:45.90 & +17:53:15.2 & 300 & 4 \\
L1257 & 23:57:31.87 & +59:39:42.1 & 140 & 5 \\
   \hline
   \end{tabular}
\footnotesize {References for distance D. (1)~\citet{elias1978}; (2)~\citet{myers1983}; (3)~\citet{bok1974};  (4)~\citet{ kawamura1998};  and (5)~\citet{snell1981}.}
 \end{table}

\section{Sources} \label{sec:sour}
\subsection{L1544}
L1544 is a well--observed molecular cloud associated 
with the Taurus dark cloud. 
There are no reported Infrared Astronomical Satellite (IRAS) sources 
associated with it~\citep{beichman1986,ward1994,doty2005}. 
 L1544 has a kinetic temperature of 9\,K with a flat uniform radial flux density in 
 inner region of radius 4800\,AU~\citep{kirk2005}, an average density of 
 3$\times$10$^{6}$ cm$^{-3}$ within 
 a radius of 500\,AU~\citep{keto2010}, 
 and extended inward motions~\citep{tafalla1998}. 
Observations of water probing the densest central region together with those of 
carbon monoxide probing the more extended cloud have given almost 
unique information about the velocity field within this dense core.
Only the unstable quasi-equilibrium Bonnor--Ebert (BE) sphere model was found to be 
consistent with molecular line observations of L1544 by~\citet{keto2015}.
The initial conditions of L1544 have been used as a testing ground 
for theories of low-mass star formation. 
For example,~\citet{whitworth2001} proposed an analytic model 
for the initial conditions of L1544. 
The model predicted the observed density profiles, the lifetimes, 
and the accretion rates of Class 0 protostars. 
The inferred infall velocities have been observed~\citep{tafalla1998,williams1999}. 
For the distance of L1544, we use the average values (140\,pc) 
from different techniques as described by~\citet{elias1978}.
  
\subsection{L1523}
L1523 is an isolated dark cloud without IRAS sources
~\citep{beichman1986} and has narrow $^{13}$CO (1--0) line widths, 
indicating cold, quiet internal physical condition~\citep{kim2002,kim2008}. 
In addition, there is CO depletion in L1523.~\citet{myers1983} used 
$^{13}$CO (1--0) and C$^{18}$O (1--0) to estimate 
a reliable distance of 140\,pc for L1523.

\subsection{B227}
B227, also called L1570, is located in a complex region of considerable extinction 
with a Galactic latitude of -0$^{\circ}$.46. 
The cloud is approaching collapse~\citep{stutz2009}. 
The optical images of B227 show an opaque, 
elongated (along the North-South direction) core surrounded 
by a more diffuse dust structure. 
There is a prominent ``ring" of H I self-absorption in B227, 
which implies that B227 is undergoing H$_{2}$ formation~\citep{zuo2018}. 
B227 is assumed to be at distances in the range between 400\,pc 
and 600\,pc~\citep{hilton1995} with the most probable distance 
accepted for various studies being 400\,pc
~\citep[see e.g.][]{bok1974,goldsmith2005,stutz2009,zuo2018}. 
For this work, we adopt the distance of 400\,pc for B227.

\subsection{L1574 and CB45}
L1574 and CB45, also named L1578, were identified as isolated dark clouds, and coincidently aligned in a linear configuration from north to south spanning about two degrees~\citep{martin1978}.
~\citet{goldsmith2005} revealed that L1574 and CB45 have H I narrow self-absorption (HINSA)  features using OH (Arecibo) and $^{13}$CO and C$^{18}$O (FCRAO). There are no nearby UV sources such as massive stars or H II regions.~\citet{kawamura1998} lumped L1574 and CB45 together, and assigned them a distance of 300\,pc.

\subsection{L1257}
L1257 is an elongated cloud from the southeast to the northwest, 
about 1$^{\circ}$ north of L1253. 
$^{13}$CO and C$^{18}$O column densities of the cloud are power-law form, 
$\rho(r) \sim r^{-1}$~\citep{arquilla1985}.
~\citet{frerking1982} found that the CO(1--0) line of L1257 has 
a possible pedestal feature and multiple components. 
~\citet{snell1981} estimated a distance of 140\,pc to L1257 by using V, (B--V) 
and MK spectral type of stars. He also determined the total mass of the cloud 
to be 35 M$_\sun$ from $^{13}$CO and H$_{2}$CO observations.

\section{Fitting Clump and velocity gradient} \label{sec:analyze}

\subsection{Identification of 2MASS Extinction Clumps with GAUSSCLUMPS} \label{sec:identi}

\begin{figure*}
\centering
\includegraphics[width=0.90\textwidth]{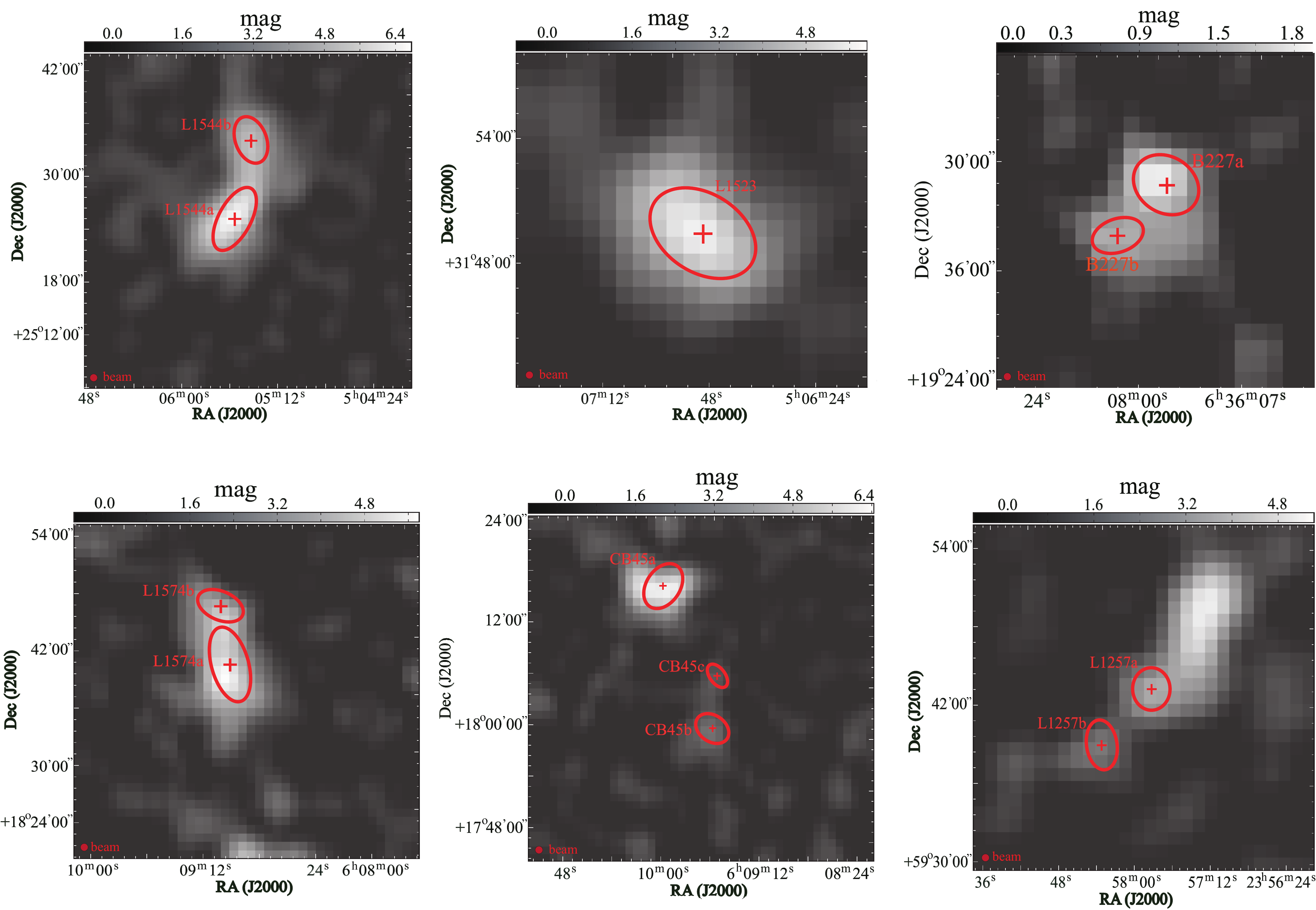} 
\caption{\footnotesize{Clumps extracted from 2MASS 
extinction maps by utilizing GAUSSCLUMPS. 
The plus symbols denote the center position of each fitted clump. 
The coordinates of each plus symbol are listed in  Table~\ref{tab:clump}.}
 \label{fig:clump1}}
\end{figure*}

Clumps appear as enhancements of column density within clouds, 
and with reasonable generality we have assume an ellipsoid shape for the clumps. 
We used the GAUSSCLUMPS~\citep{stutzki1990} code 
in the CUPID~\citep{berr2007} package to identify clumps. 
This package is part of Starlink software\footnote{http://starlink.jach.hawaii.edu/starlink/}. 
GAUSSCLUMPS searches for an ellipsoidal Gaussian density profile 
around the density peaks,
and subtracts the fitted Gaussian ellipsoid from the data. 
It then iterates on the ``clump removed data'' until terminating criteria 
({\rm\bf MAXSKIP}~$\leqslant$~50, {\rm\bf THRESH}~$\geqslant$~5, 
{\rm\bf NPAD}~$\leqslant$~50, {\rm\bf MINPIX}~$\geqslant$~25,  
{\rm\bf MAXNF}~$\leqslant$~200) are reached (Table~\ref{tab:para}). 
Following the instructions of CUPID, 
we first subtracted the background with the FINDBACK procedure. 
After the background subtraction, 
we used the GAUSSCLUMPS to fit Gaussian components in the 2MASS extinction map. 
In the fitting process, ``usable'' clumps are those that are well fitted with peak intensity 
five times higher than the Root Mean Square (rms) noise following~\citet{qian2012}.
A sample of 12 fitted clumps was thus obtained. 
Figure~\ref{fig:clump1} shows the fitted clumps.
\begin{table}[!htbp]
\caption{Parameters used in the GAUSSCLUMPS method to identify clumps in the extinction map. \label{tab:para}}
 \vspace*{0.5ex}
\begin{tabular}{l c c c c c c c c c c c}
\hline
\hline
Parameter & {} & {} & {} & {} & {} & {} & {} & {} & {} & {} &Value \\\hline

WWIDTH & {} & {} & {} & {} & {} & {} & {} & {} & {} & {}  & 2 \\\hline

WMIN     & {} & {} & {} & {} & {} & {} & {} & {} & {} & {} & 0.01  \\\hline

MAXSKIP & {} & {} & {} & {} & {} & {} & {} & {} & {} & {} & 50 \\\hline

THRESH & {} & {} & {} & {} & {} & {} & {} & {} & {} & {} & 5 \\\hline

NPAD & {} & {} & {} & {} & {} & {} & {} & {} & {} & {} & 50  \\\hline

MAXBAD & {} & {} & {} & {} & {} & {} & {} & {} & {} & {} & 0.05 \\\hline

VELORES & {} & {} & {} & {} & {} & {} & {} & {} & {} & {}  & 2 \\\hline

MODELLIM  & {} & {} & {} & {} & {} & {} & {} & {} & {} & {} & 0.05  \\\hline

MINPIX    & {} & {} & {} & {} & {} & {} & {} & {} & {} & {} & 25  \\\hline

FWHMBEAM & {} & {} & {} & {} & {} & {} & {} & {} & {} & {} & 2 \\\hline

MAXCLUMPS & {} & {} & {} & {} & {} & {} & {} & {} & {} & {} & 2147483647\\\hline

MAXNF     & {} & {} & {} & {} & {} & {} & {} & {} & {} & {} & 200\\\hline
\end{tabular}

\footnotesize {\rm\bf Notes.}
{\rm\bf WWIDTH} is the ratio of the width of the
weighting function, which is a Gaussian function, to that of the initial guessed Gaussian function.

{\rm\bf WMIN}: specifies the minimum weight. Pixels with weight less than this value are not included in the fitting process.

{\rm\bf MAXSKIP}: The iterative fitting process is terminated if more than ``{\rm\bf MAXSKIP}"  consecutive clumps cannot be fitted.

{\rm\bf THRESH}: gives the minimum peak amplitude of clumps to be fitted by GAUSSCLUMPS. The supplied value of ``{\rm\bf THRESH}" is multiplied by the {\rm\bf rms} noise level before being used.

{\rm\bf NPAD}: The algorithm will terminate when ``{\rm\bf NPAD}" consecutive clumps have been fitted with peak values less than the threshold value specified by the ``{\rm\bf THRESH}" parameter (From the source code CUPID GaussClumps.c, one can see that the algorithm will do the same thing when ``{\rm\bf NPAD}" consecutive clumps have pixels fewer than ``{\rm\bf MINPIX}").

{\rm\bf MAXBAD}: is the maximum fraction of bad pixels that may be included in a clump. Clumps will be excluded if they contain more bad pixels than this value.

{\rm\bf VELORES}: is the velocity resolution of the instrument in channels.

{\rm\bf MODELLIM}: is the model values that are treated as zero below ModelLim times the {\rm\bf rms} noise.

{\rm\bf MINPIX}: is the lowest number of pixel contained in a clump.

{\rm\bf FWHMBEAM}: is the FWHM of the instrument beam in pixels.

{\rm\bf MAXCLUMPS}: is an upper limit to the number of clumps to be fitted. It is set to be a very large number so that this parameter does not take effect.

{\rm\bf MAXNF}: is the maximum number of the allowed evaluations of the objective function when fitting an individual clump. Herein, it is just set to be a very large number to guarantee that all the clumps are fitted.
\end{table} 

For fitted ellipsoids, the fitted clump radius can be defined as the geometrical mean of the semi-major ($R_{major}$) and semi-minor ($R_{minor}$) axes: 
\begin{equation}
R = (R_{major}R_{minor})^{1/2} \label{ident1}.
\end{equation}

GAUSSCLUMPS outputs the total intensity, $A$${_v^{tot}}$, 
which depends on the sum of all the pixels in each fitted clump. 
The fitted clump mass can be calculated by the following relation: 
\begin{equation}
M = R^{2}\mu \beta_{v} A{_v^{tot}}  \label{ident2}.
\end{equation}
$\mu$ denotes the mean molecular weight corrected for the Helium (He) abundance.  
We take the Helium abundance to be 
[H$_{e}$]/[H$_{2}$] = 2/9 with $\mu = 2.88$. 
$\beta$$_{v}$ $\simeq$ 1.9$\times$10$^{21}$cm$^{-2}$mag$^{-1}$ is 
the ratio ($N$(H$_{\uppercase\expandafter{\romannumeral1}}$)+ 
2$N$(H$_{2}$))/A$_{v}$~\citep[]{bohlin1978}. 
Table~\ref{tab:clump} gives the parameters of the fitted clumps. 

\begin{figure}[!htbp]
\centering
\includegraphics[width=0.35\textwidth]{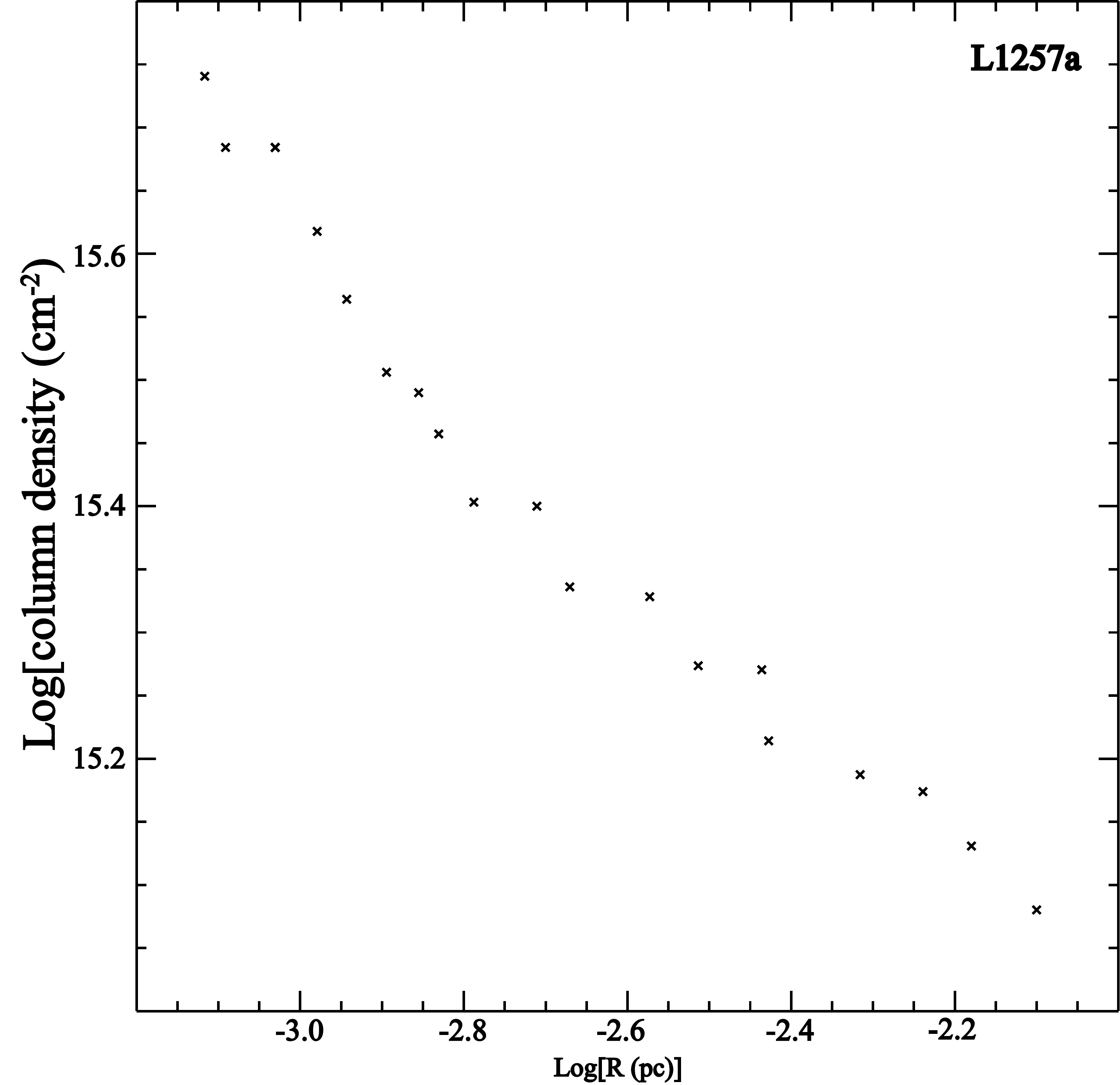}
\caption{The distribution of the C$^{18}$O column density in clump L1527a.\label{fig:column}}
\end{figure}

The density profile is expected to have a power law form with a central core
when the cloud is dominated by gravity~\citep[e.g.][]{ballesteros2011,chen2018}. 
The Figure 1 of~\citet{keto2010} shows that 
density profiles of starless cores L1544 have an outer region 
where the density scales as $\rho{}\propto r^{-2}$ 
and an inner region ($\simeq{}3.5^{''}$) 
with constant density of 2$\times$10$^{7}$ cm$^{-3}$. 
The sizes of our fitted L1544 clumps are much bigger than their inner regions. 
Figure~\ref{fig:column} presents the C$^{18}$O column density 
distribution in clump L1257a. 
We conclude that density increases from the outer region to the inner 
region in our fitted clumps. 
For the present work we adopt a power-law profile density 
$\rho{}\propto{}r^{-1.6}$~\citep{bonnor1956}, 
which is reasonably close to various models~\citep[e.g.][]{keto2010,di2013,keto2015}.

\subsection{Line Fitting with CLASS} \label{sec:line}
We used the CLASS software in 
Gildas\footnote{http://www.iram.fr/IRAMFR/GILDAS}
to fit $^{13}$CO(1--0) and C$^{18}$O(1--0)  spectral lines. 
Each spectrum was fit with a single Gaussian profile. 
There are three fit parameters in this process: 
the peak antenna temperature T$_{A}^{*}$,  
the line of sight velocity v$_{LSR}$, and the line width (full width at half-maximum, FWHM) $\Delta{}v$.  
The statistical uncertainties of each parameter were computed using expressions 
for the uncertainties obtained from~\citet{landman1982}:
\begin{equation}
\sigma_{T_A^{*}} = 1.41(\frac{\delta_v} {\Delta{}v})^{1/2} \sigma_{rms},
\end{equation}
\begin{equation}
\sigma_{v_{LSR}} = 0.69(\delta_v{}\Delta{}v)^{1/2} \frac{\sigma_{rms}} {T_A^{*}} \label{equ1}.
\end{equation}
$\delta_v$ is the velocity resolution and $\sigma_{rms}$ is the rms  noise in the spectrum.

\begin{figure*}
\centering
\includegraphics[width=0.70\textwidth]{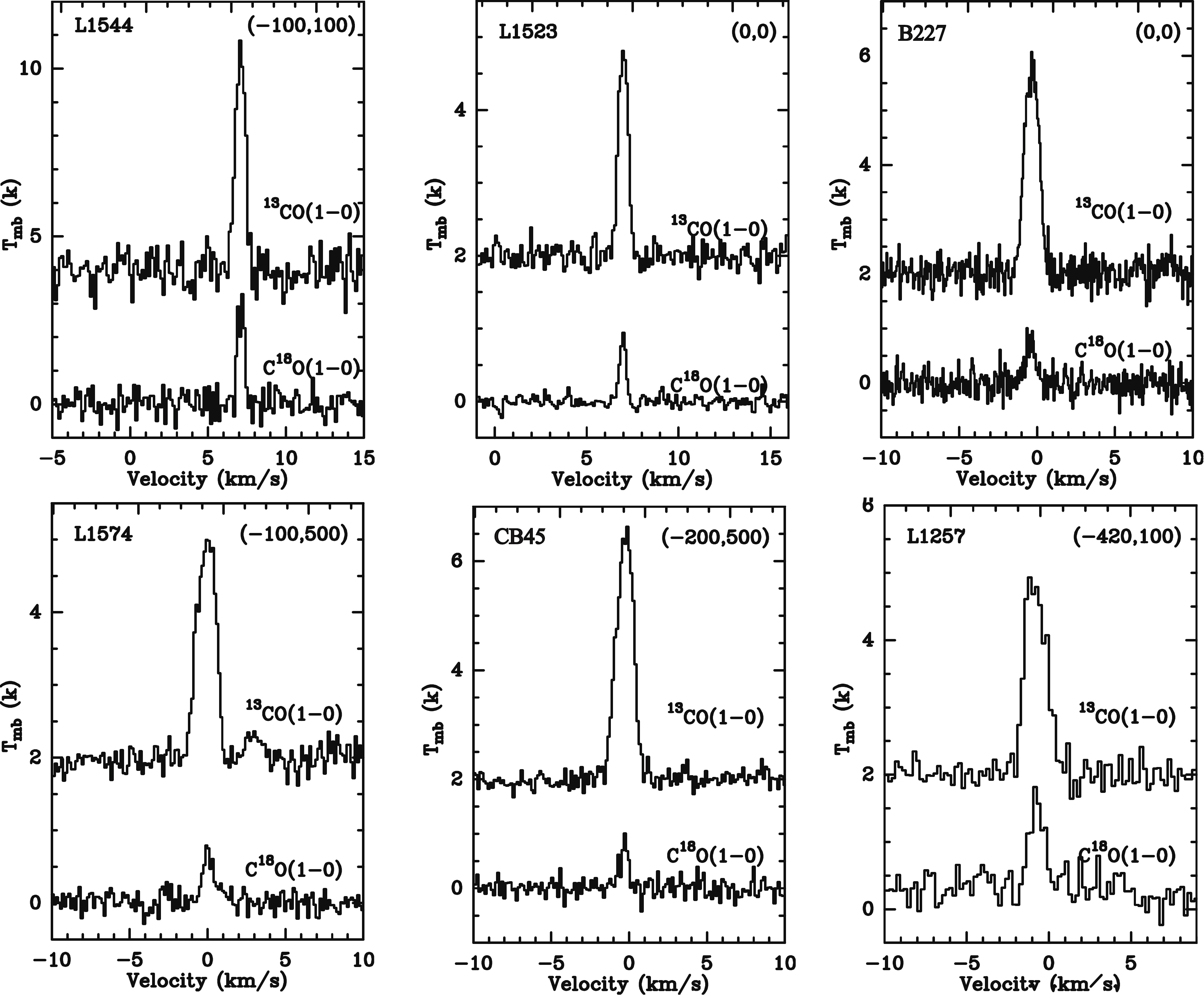}
 \caption{\footnotesize{Spectra of $^{13}$CO and C$^{18}$O in the six clouds included in this study. L1523 and B227 spectra are at the nominal central position of each cloud, while the offsets for L1544, L1574, CB45 and L1257 are in seconds of arc offset relative to the central position. The central positions are listed in Table~\ref{tab:sources}.} \label{fig:spectra}}
\end{figure*}

We chose spectral lines to fit the velocity field of every fitted clump in three steps. 
First, we selected $^{13}$CO(1--0) and C$^{18}$O(1--0) spectral lines 
with peak intensity greater than three times of the rms noise and fitting errors. 
Figure~\ref{fig:spectra} presents the two selected lines in the 6 molecular clouds. 
Second, these two lines in first step were located within the fitted 2MASS 
extinction image of the fitted clump. 
Third, the lines identified in the second step were then used to fit the velocity field.
Table~\ref{tab:width} presents the fitting results for all 12 identified clumps. 
The optical depth $\tau$ was calculated assuming that 
$^{13}$CO(1--0) and C$^{18}$O are optically thick and thin, respectively, 
and that the temperature of the fitted clumps is 10 K.

\subsection{Velocity Gradient Fitting} \label{sec:gradient}

Rigid-body rotation in an molecular cloud produces a linear gradient, 
$\mathbf\nabla{}v_\mathrm{LSR}$. 
In the local standard of rest (LSR) velocity field, 
the linear gradient is in a direction perpendicular to the rotation axis. 
We measured the velocity gradients following the method 
described in~\citet{goodman1993} fitting the function
\begin{equation}
v_{LSR} = v_0 + c_{1}{}\Delta{}\alpha + c_{2}{}\Delta{}\delta  
\label{equ2}.
\end{equation}
Here $v_{LSR}$ represents an intensity weighted average velocity 
along the line of sight and $v_{0}$ is the systematic clump velocity, 
$\Delta{}\alpha$ and $\Delta{}\delta$ are the offsets from the center positions 
(see Table~\ref{tab:sources}) of the $^{13}$CO and C$^{18}$O images 
in the right ascension and declination in radians, respectively,  
and $c_{1}$ and $c_{2}$ are the projections of the gradient per radian onto 
the $\alpha$ and $\delta$ axes, respectively. 
The magnitude of the velocity gradient is defined by
\begin{equation}
{\cal{}G}= |\mathbf\nabla
v_\mathrm{LSR}|=\frac{(c_{1}^{2}+c_{2}^{2})^{1/2}}{D},
\end{equation}
where $D$ is the distance to the object.
Its direction (i.e., the direction of the increasing velocity, measured
east of north) is given by
\begin{equation}
\theta_{\cal G}= \tan{}\frac{c_{1}}{c_{2}}.
\end{equation}
We performed a least-squares fit to the Equation (\ref{equ2}). 
Each value of $v_{LSR}$ was weighted by 
$\sigma_{v_{LSR}}^{-2}$(Equation~(\ref{equ1})).
The magnitude of the gradient ($\cal{}G$), its direction ($\theta_{\cal{}G}$), 
and their errors were calculated based on the results 
($c_{1}$ and $c_{2}$) of the least-squares fit.

When a fitted clump has at least nine spatial pixels, 
its velocity gradient can be reliably fitted~\citep{goodman1993}.
This applies to our 12 2MASS extinction clumps. 
All of the 12 clumps had well-fitted $^{13}$CO, while 11 clumps, 
with the exception of B227b, had well-fitted C$^{18}$O lines. 
Beam smearing (inadequate spatial resolution) can
possibly affect the fitting of velocity gradients.
To test this, we have smoothed the spectral images of L1544 and B227, 
and then fitted the velocity gradient. 
We found that on average the finite beam size
reduces the velocity gradient by 7.8\%.
Our sources are relatively near so that beam smearing should
have only a minor impact on our results.
For similar reasons, beam smearing was not discussed in the previous work on cloud rotation
~\citep[]{clark1982,arquilla1986,goodman1993,kane1997,caselli2002,
pirogov2003,redman2004,shinnaga2004,chen2007,chen2009,li2012,tatematsu2016}.     

\begin{figure}[!htbp]
\centering
\includegraphics[width=0.35\textwidth]{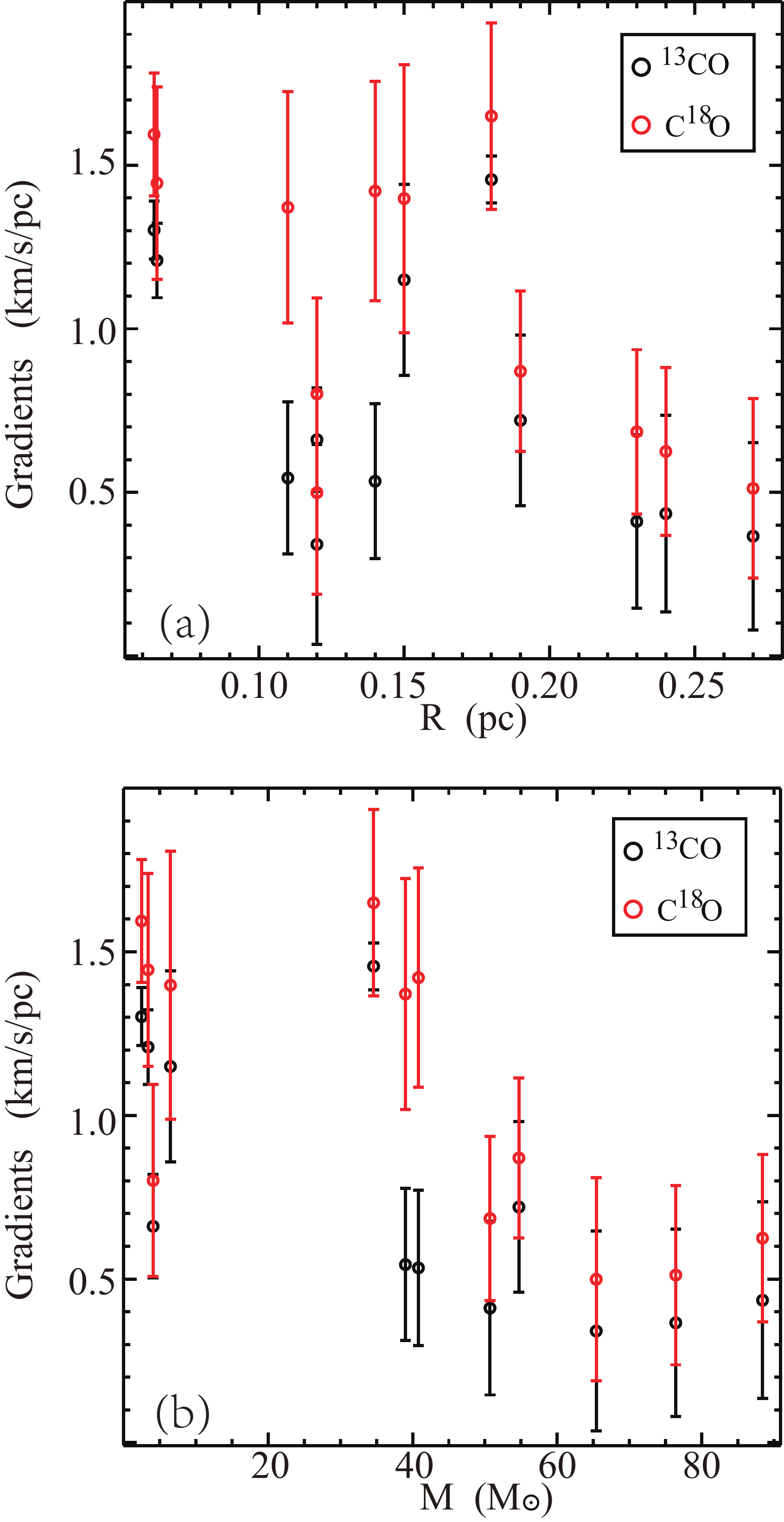}
\caption{Velocity gradients correlated with (a) clump size, $R$ and (b) clump mass, $M$.\label{fig:gradient2}}
\end{figure}

\begin{figure}[!htbp]
\centering
\includegraphics[width=0.35\textwidth]{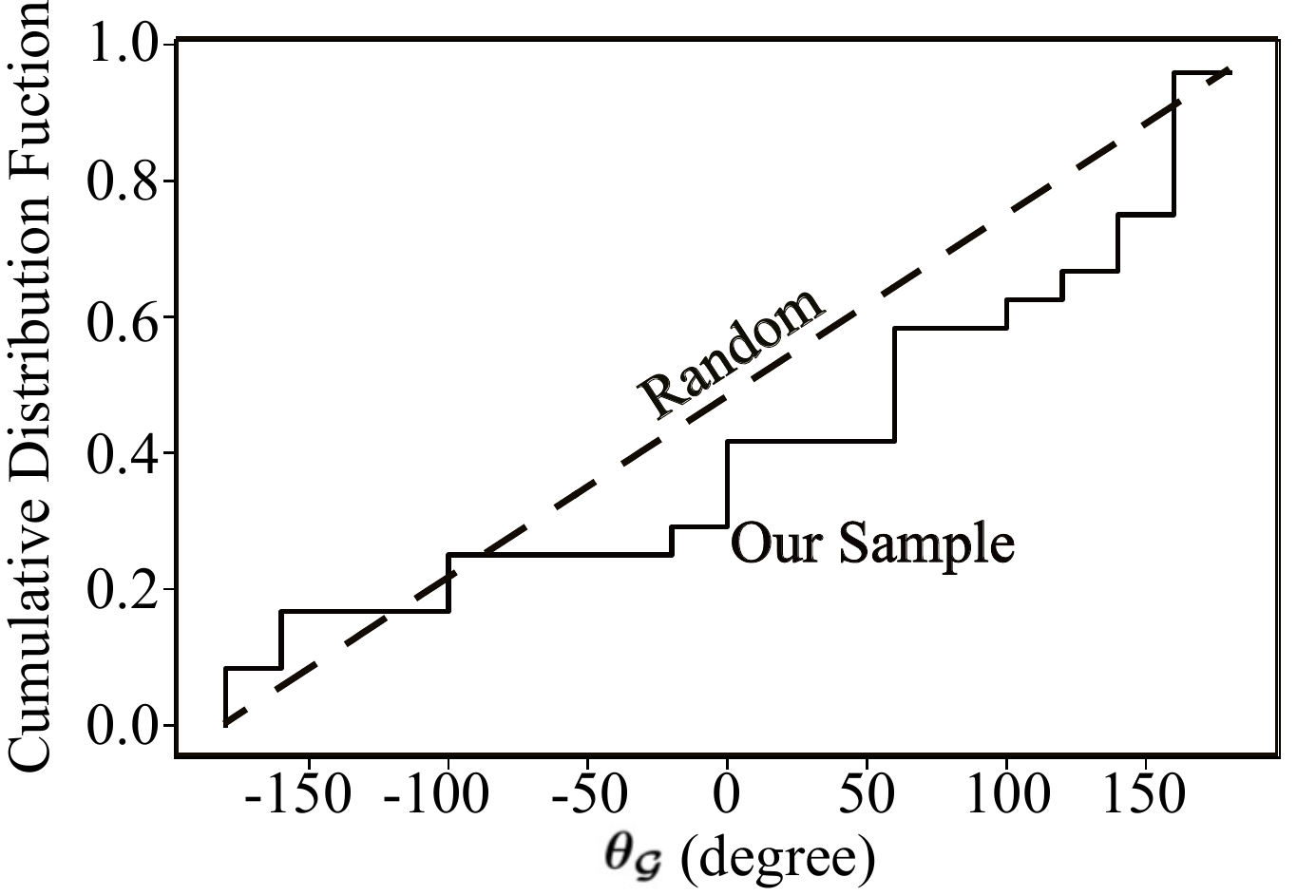}
\caption{The cumulative distribution function of velocity gradient 
direction ($\theta_{\cal G}$). 
\label{fig:angle_hist}}
\end{figure}

The fitted $\cal G$ and $\theta_{\cal G}$ are listed in Table~\ref{tab:gradient1} 
and plotted in Figure~\ref{fig:gradient2}. 
The C$^{18}$O gradients are consistently larger than those of  $^{13}$CO. 
One possibility is that C$^{18}$O is better in tracing our fitted clumps. 
We also found that $\cal G$  increases as $M$ decreases, 
and that there is no obvious correlation between $\cal G$ and $R$. 
Figure~\ref{fig:angle_hist} shows the cumulative distribution function 
of  $\theta_{\cal G}$,
which is largely consistent with a random orientation.  

\section{Specific Angular Momentum Analysis} \label{sec:rotation}

\begin{figure}[!htbp]
\centering
\includegraphics[width=0.35\textwidth]{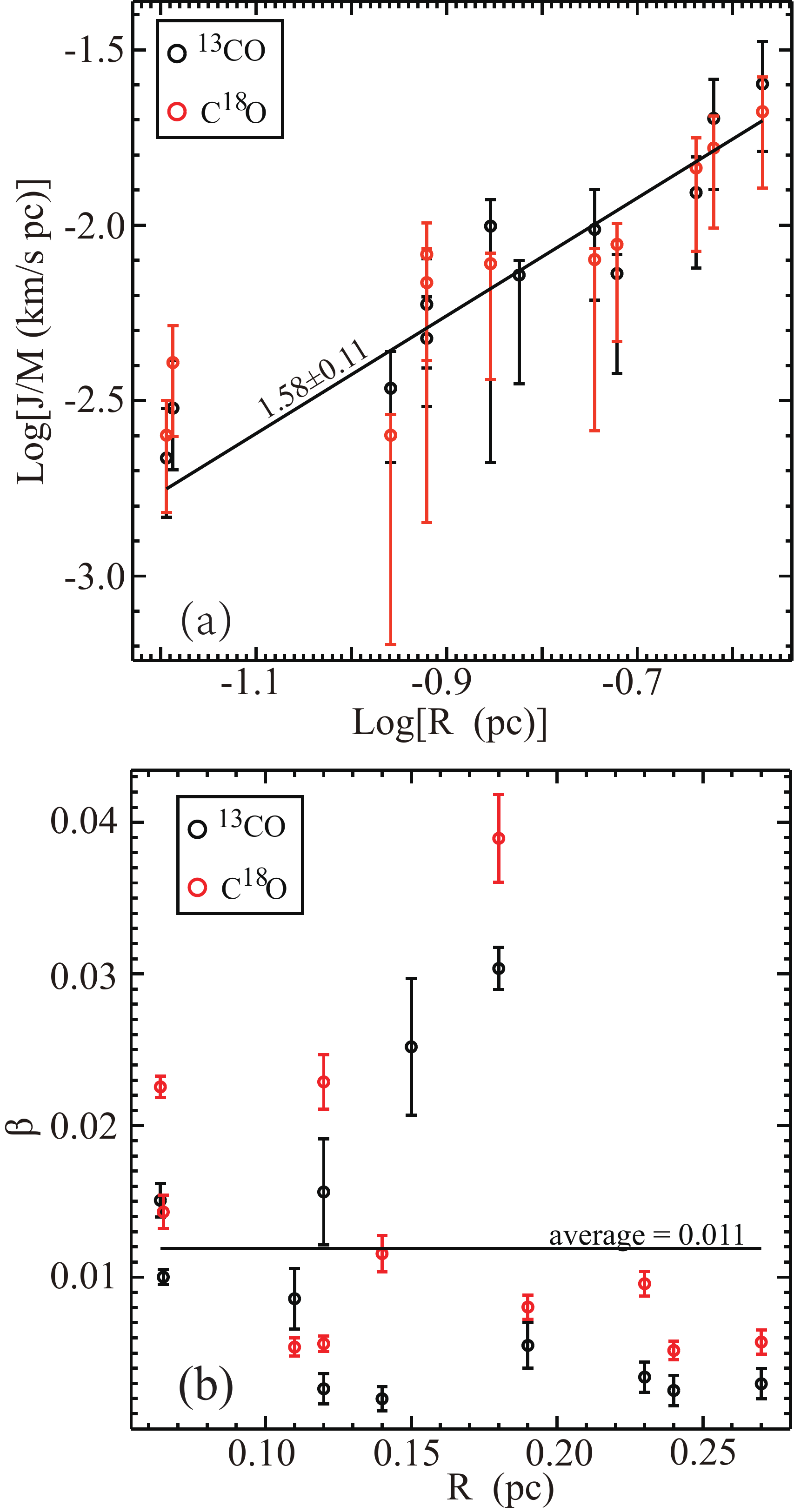}
\caption{(a) $J/M$ and (b) $\beta$ correlated with clump size, $R$. The best power-law relation is obtained between $J/M$ and $R$.\label{fig:size}}
\end{figure}

\subsection{Specific Angular Momentum}\label{sec:sam}
If molecular clouds are rotating, their present--day angular momentum could shed light on their evolution.  
The angular momentum per unit mass, $j$, is often used to compare the angular momenta 
at different regions with comparable mass. 
The specific angular momentum is given by
\begin{equation}
j = \frac{J}{M} =\frac{I\omega}{M}  = p{}\omega R^{2},
\end{equation}
where $I = pMR^{2}$ denotes the moment of inertia for a fitted clump 
with power-law density distribution ($\rho~\propto~r^{-\cal{}A}$). 
The correction factor ($p$) to the momentum of inertia due 
to density distribution can be expressed as 
\begin{equation}
p = \frac{2(3-{\cal{}A})}{3(5-{\cal{}A})}.
\end{equation}
The detailed derivation of $p$ is shown in the Appendix A. 
$\omega={\cal{}G}/\sin i$ is  the angular velocity of the fitted clump, 
where $i$ represents the inclination of $\omega$ to the line of sight. 
Assuming $\sin i = 1$, we find
\begin{equation}
J/M =  \frac{2(3-{\cal{}A})}{3(5-{\cal{}A})} R^{2}\label{equ3}.
\end{equation}

The slope of the power-law density distribution 
 was described in section~\ref{sec:identi}.
The assumption of $\cal{}A$ = 1.6 results in a 70\% smaller specific 
angular momentum ($J/M$) than that of a uniform sphere. 
The calculated $J/M$ of our fitted clumps is presented in 
Table~\ref{tab:momentum} 
and plotted in Figure~\ref{fig:size}(a). 
The fitted power--law relation between $J/M$ and $R$ is 
$J/M~\propto~R^{1.58~\pm~0.11}$. 
Although marginally lower than what was found (1.6) 
in~\citet{goodman1993}, our result generally agree with theirs.

Since we adopted $\sin i = 1$, but as we do not know the inclination of 
a given cloud along our line of sight, our determined $\omega$ could 
underestimate the true value (${\cal{}G}/\sin i$).  
Therefore, our measurement of $J/M$ is likely to be an underestimate, 
especially in a statistical sense.
 
\subsection{Compare Rotational Energy with Gravitational Energy} \label{sec:ratio}
In this section, we compare rotational energy with gravitational energy. 
Their ratio is defined as: 
\begin{equation}
\beta =  \frac{E_{r}}{E_{g}},
\end{equation}
where $E_{r} = \frac{1}{2}I\omega^{2} = \frac{1}{2}pMR^{2}\omega^{2}$  
denotes the rotational energy, and 
\begin{equation}
E_{g} = -\frac{3}{5}\lambda{}\gamma{}\frac{GM^2}{R}
\end{equation}
is the gravitational energy of the mass $M$ within a radius $R$. 
The enhancement factor to the potential energy due to a power-law density profile is 
$\lambda = (1-{\cal{}A}/3)/(1-2{\cal{}A}/5)$~\citep{di2013}.
$\gamma$ represents the enhancement factor of gravitational potential 
due to deviation from spherical symmetry and can be calculated~\citep{di2013} as
\begin{gather}
\gamma = \frac{\arcsin{\sqrt{1-f^2}}}{\sqrt{1-f^2}}, \notag \\ \label{prolate}
f = \frac{2}{\pi}f_{obs}\mathcal{F}_{1}(0.5,0.5,-0.5,1.5,1,1-f^2_{obs}) ,
\end{gather}
where $f_{obs} = R_{minor} / R_{major}$ is the observed axis ratio of clumps,  
$\mathcal{F}_{1}$ denotes the Appell hypergeometric function in its first form.
The $\gamma$ value for each fitted clump is in Table~\ref{tab:clump}. 

The assumption of $\cal{}A$ = 1.6 results in 70\% smaller E$_{r}$ 
and 31\% bigger E$_{g}$, 
which add up to a factor of 2 lower $\beta$ than that of a uniform sphere.
For the inclination $i$, $\beta$ is underestimated for the same reason 
as discussed above for $J/M$. 
The calculated $\beta$ values can be found in Table~\ref{tab:momentum} 
and were plotted in Figure~\ref{fig:size}(b). 
$\beta$ does not vary with clump size $R$, 
and may be a constant independent of $R$. 
This conclusion is consistent with that of~\citet{goodman1993}.

Rotational energy can be important for supporting dark cloud envelops 
against gravitational collapse~\citep[e.g.][]{field1978,boss1999,dib2010}.
Rotation can prevent gravitational collapse when $\beta$~$\geqslant$~0.25.
It is difficult to determine the status of cores when $\beta$~\textless~0.1.
For our clumps, $\beta$ ranges from 0.0012 to 0.018.
These values are too small to prevent gravitational collapse of the clumps.

\section{Discussion} \label{sec:discuss}
\subsection{Comparison of $^{13}$CO and C$^{18}$O Gradient Directions}
\begin{figure}[!htbp]
\centering
\includegraphics[width=0.30\textwidth]{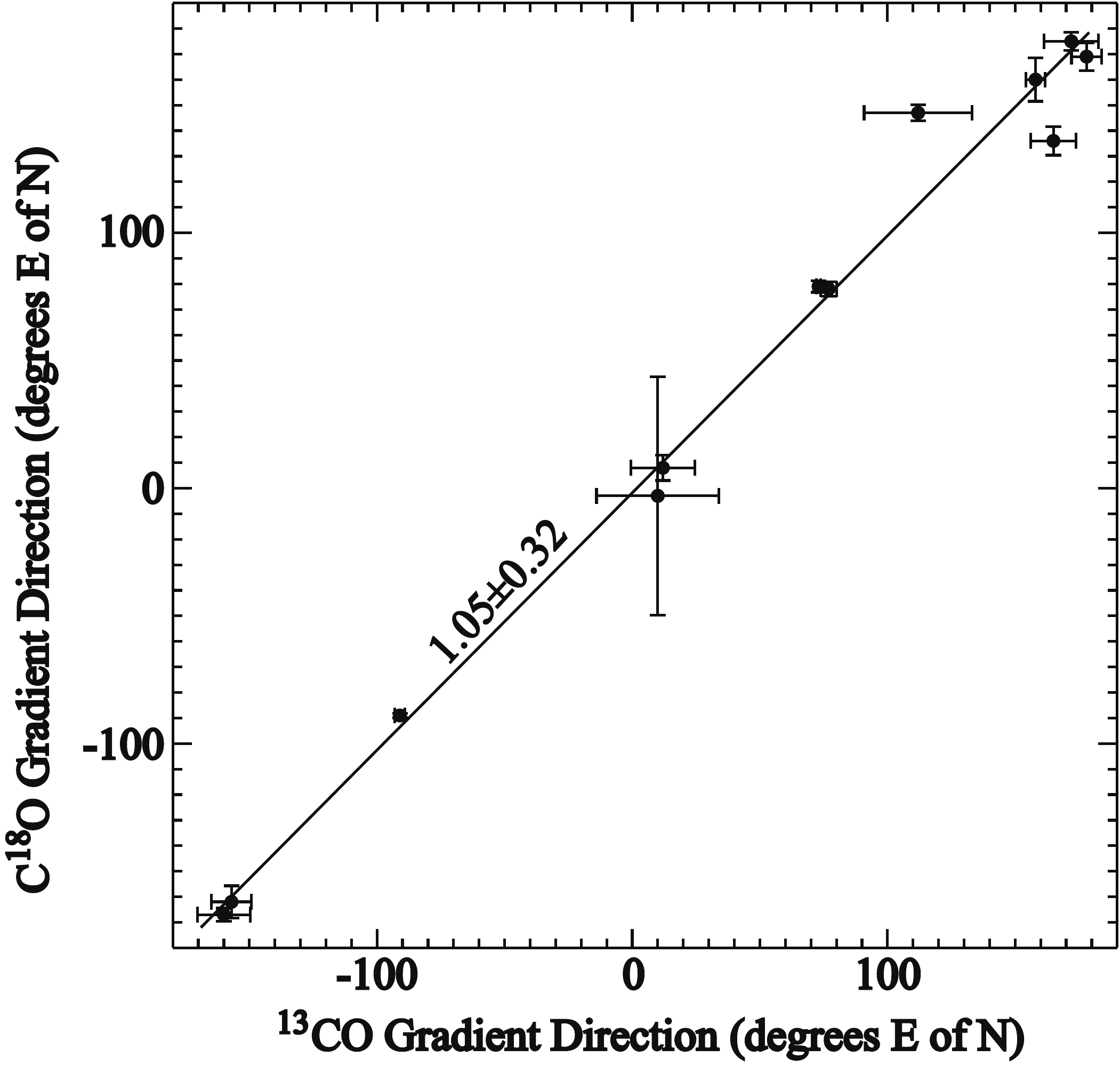}
\caption{Comparison of gradient directions found using $^{13}$CO and C$^{18}$O.\label{fig:angle}}
\end{figure}

Figure~\ref{fig:angle} shows the correlation between $^{13}$CO 
and C$^{18}$O velocity gradient directions. 
We found a slope 1.05~$\pm$~0.32 from the least squares linear fit, 
with a confidence level of 75\%.
This significant correlation indicates that $^{13}$CO and C$^{18}$O 
are tracing the same clump.
~\citet{myers1991}proposed that there is a good correlation 
between position angles in different tracers for the same clump.
This is confirmed by our results. 

\subsection{The Alignment Between Clump Elongation and Rotation Direction} 
\label{sec:shape}

\begin{figure}[!htbp]
\centering
\includegraphics[width=0.30\textwidth]{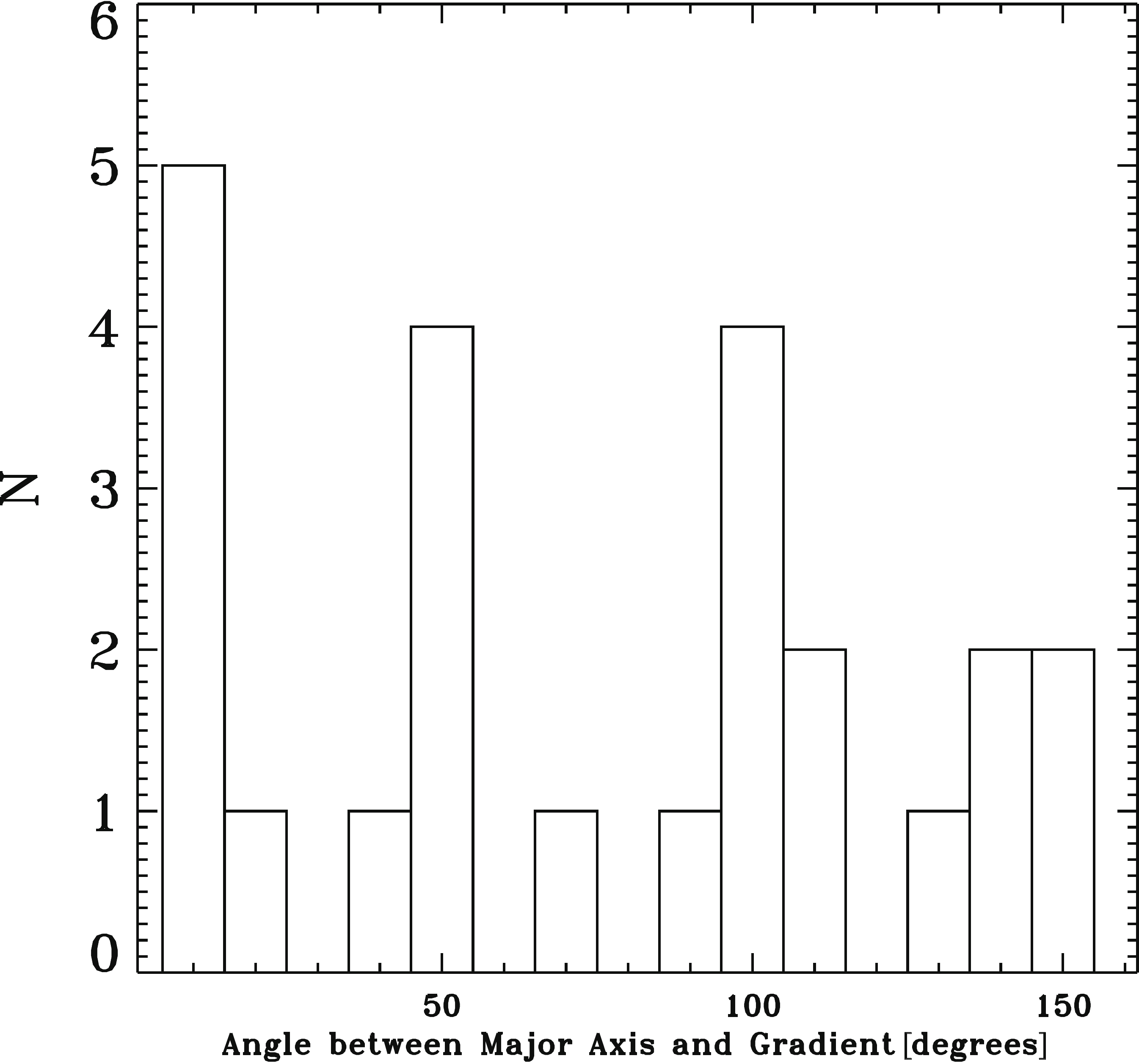}
\caption{Distribution of angles between the major axis of each clump and 
the direction of its velocity gradient. Both $^{13}$CO and C$^{18}$O 
gradient directions of each clump are included. \label{fig:distribute}}
\end{figure}

\begin{figure}[!htbp]
\centering
\includegraphics[width=0.30\textwidth]{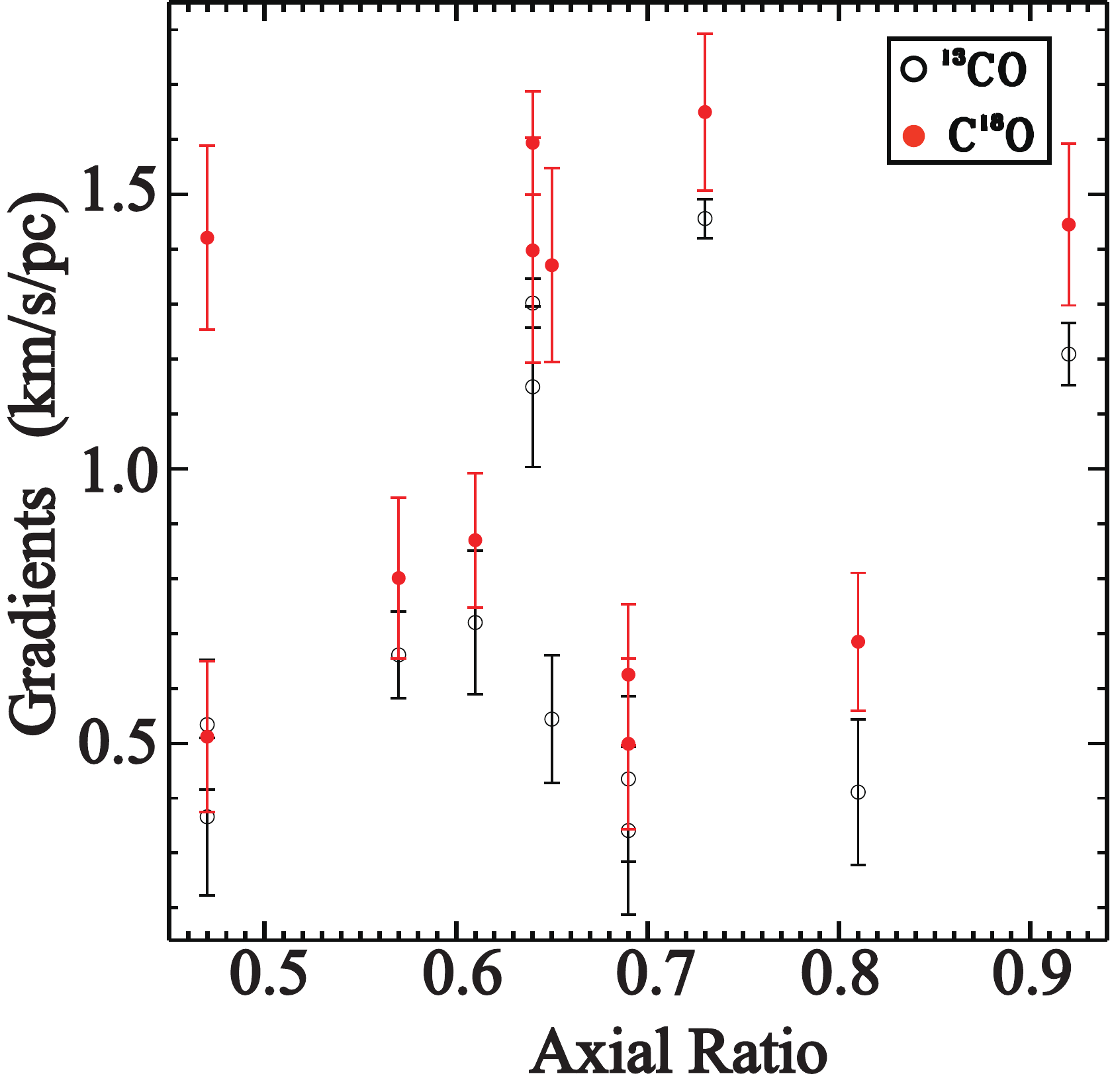}
\caption{Velocity gradients correlation with the axial ratio of clumps.\label{fig:axis}} 
\end{figure}

Clump rotation sufficient to produce appreciable velocity gradients, 
can also lead to clumps being flattened along the axis of rotation. 
In our sample, the directions of velocity gradients are often different 
from the position angle of the long axis of a clump. 
The angular difference is likely random as shown in Figure~\ref{fig:distribute}. 
We note that there is no correlation between the clump elongation and the rotation direction. 
Figure~\ref{fig:axis} displays $\cal{}G$ as a function of clump axial ratio, 
with no correlation being evident. 

\subsection{J/M and $\beta$ in This Study and Previous Studies}

\begin{figure*}[!htbp]
\centering
\includegraphics[width=0.75\textwidth]{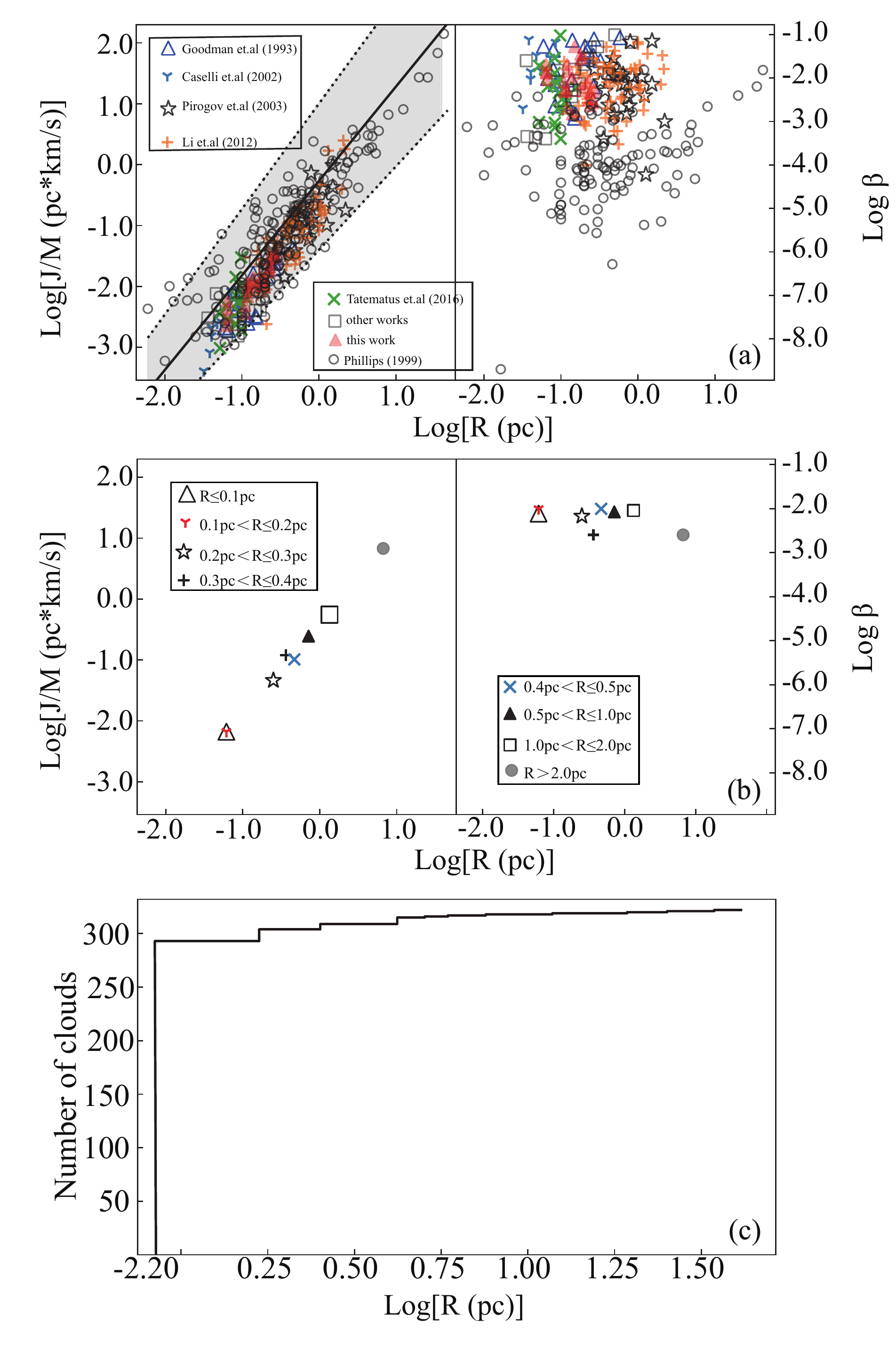}
\caption{(a) Specific angular momentum ($J/M$) and the ratio of rotational to 
gravitational energy ($\beta$) as a function of $R$ for the clumps in the present study 
together with those previously observed. 
The left panel shows the best--fit slope, 1.52,  for log$J/M$ vs log $R$. 
The dotted lines denote the best--fit slope with +/-0.13 uncertainties. 
The right panel displays $\beta$ vs $R$. 
(b) Averages $J/M$ and $\beta$ as a function of $R$.
(c) Distribution of combined samples in this and previous studies as function of $R$. 
 \label{fig:specific}}
\end{figure*}

We compare the $J/M$ and $\beta$ values in our sample with those found in previous studies. 
Figure~\ref{fig:specific}(a) compares $J/M$ and $\beta$. 
There exists a power-law relation between $J/M$ and clump size $R$ 
for the entire ensemble of measurements, given by $J/M~\propto~R^{1.52~\pm~0.13}$. 
The value of the slope, 1.52, and its 1 sigma uncertainties, +/-0.13, 
form a well-defined region (the shaded region of Figure~\ref{fig:specific}(a)). 
The value 1.52 is consistent with value of 1.58, 
the best--fit value for the clumps studied here . 
$\beta$ shows a large scatter, but the small value of $\beta$ indicates 
that the rotational energy is a small fraction of the gravitational energy. 
Thus, the observed rotation alone cannot stop the gravitational collapse of 
the clumps found in molecular clouds.   

Figure~\ref{fig:specific}(c) plots the distribution of all clumps with $J/M$ and $R$ binned. 
The average values of $J/M$ and $\beta$ were calculated for each bin. 
We found that the binned  $J/M$ and $\beta$ has similar trend as 
Figure~\ref{fig:specific}(a) shown.
The power-law relation between $J/M$ and $R$ applies over a range 
of spatial scales (0.006 -- 42.0\,pc). 
Due to the reduced scatter, we see that there is no evident relationship 
between $\beta$ and $R$.

\section{Summary} \label{sec:sum}
 We have mapped $^{13}$CO and C$^{18}$O $J$ = 1--0 emission from 6 molecular clouds. 
 The rotation properties of the identified clumps were studied based on the velocity gradients. 
Our results are summarized as follows.

1. We used GAUSSCLUMPS to identify clumps with the 2MASS extinction images 
and identified 12 clumps in 6 molecular clouds.
All clumps, except for B227b in C$^{18}$O, 
can be well fitted by single Gaussians, 
yielding velocity gradients between 0.34 and 1.46\,km\,s$^{-1}$\,pc$^{-1}$ 
for $^{13}$CO and between 0.50 and 1.65\,km\,s$^{-1}$\,pc$^{-1}$ 
for C$^{18}$O.

2. The specific angular momentum ($J/M$) and the ratio 
between rotation energy and gravitational energy ($\beta$) 
were calculated from the measured velocity gradients. 
We adopted a more realistic assumption 
of a power-law density profile $\rho~\propto~r^{-1.6}$, 
resulting in 30\% lower $J/M$ and 50\% lower $\beta$ than 
those of the more commonly assumed spheres of uniform densities. 

3. The calculated $J/M$ ranges from 0.0022 to 0.025\,pc\,km\,s$^{-1}$ 
for $^{13}$CO,
and from 0.0025 to 0.021\,pc\,km\,s$^{-1}$ for C$^{18}$O. 
This indicates that $J/M$ does not have significant evolution in the material 
traced by  $^{13}$CO compared to that traced by C$^{18}$O in dense clumps. 
There is a  power-law relation between the $J/M$ and the clump size ($R$), 
\textbf{$J/M~\propto~R^{1.58~\pm~0.11}$}, 
which is in general agreement with that found in~\citet{goodman1993}. 
 
4. The calculated $\beta$ ranges from 0.0012 to 0.018 and is independent of $R$. 
The small value of $\beta$ indicates that rotation alone cannot support 
the clumps against gravitational collapse.

5. The direction of clump elongation does not correlate 
with the direction of the velocity gradient. 
 
6. We assembled the $J/M$ and $\beta$ measurements from literatures and 
studied their relation to $R$, along with our results. 
For the combined data set, $J/M$ also increases monotonically with 
$R$ ($J/M~\propto~R^{1.52~\pm~0.13}$). 
The slope value of 1.52 is consistent with that of our sample alone (1.58).

7. The binned average $J/M$ and $\beta$ are well--correlated with $R$, 
and exhibit a similar trend as described above.
We found that the power--law relation between $J/M$ and $R$ applies 
over a range of spatial scales of 0.006 to 42\,pc. 

\acknowledgments
We thank the referee for his constructive suggestions.
This work is supported by the National Natural Science Foundation of China grant 
No.\ 11988101, No.\ 11725313, No.\ 11721303, 
the International Partnership Program of Chinese Academy of Sciences grant 
No.\ 114A11KYSB20160008, and the National Key R\&D Program of China 
No.\ 2016YFA0400702. This work was carried out in part at 
the Jet Propulsion Laboratory, California Institute of Technology, 
under a contract with the National Aeronautics 
and Space Administration (80NM0018D0004).

\setcounter{figure}{0}
\renewcommand{\thefigure}{A\arabic{figure}}
\appendix
\section{the Moment of Inertia  Derivation}
\begin{figure}[ht!]
\centering
\includegraphics[width=0.35\textwidth]{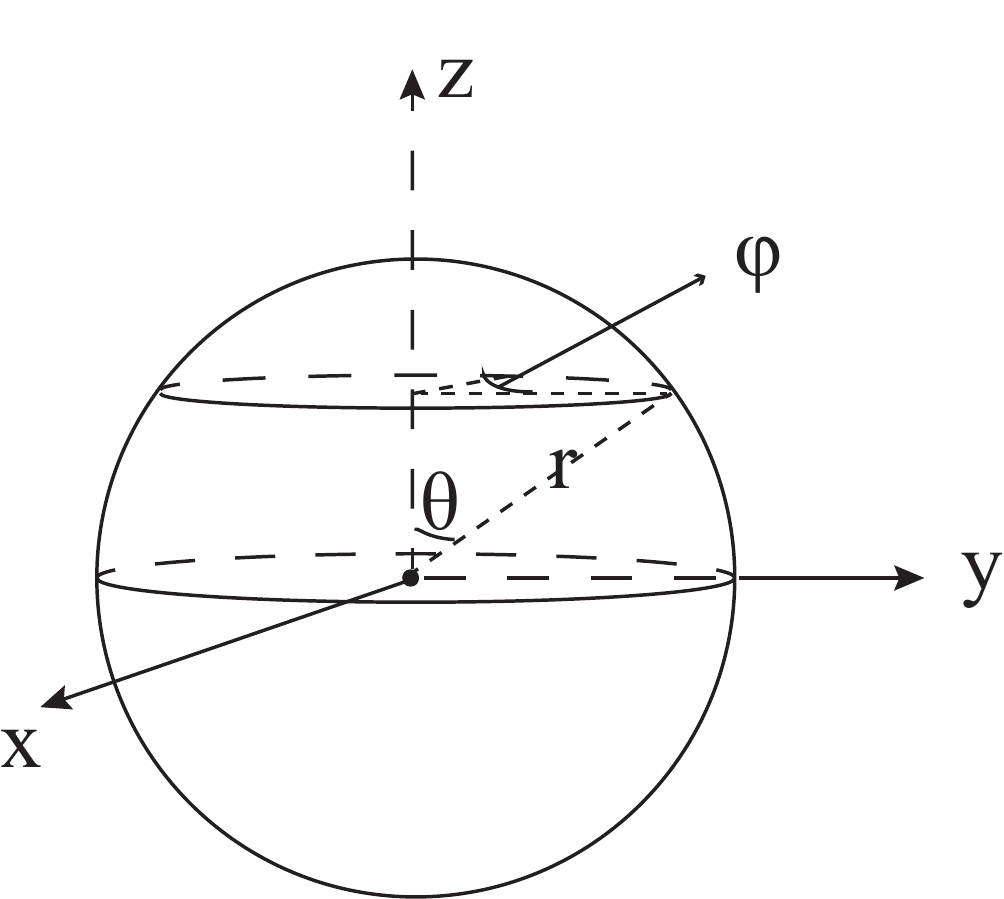}
\caption{The point (0, 0, 0) located in the center of the sphere. 
\label{fig:sphere}}
\end{figure}

The equation of a sphere (figure~\ref{fig:sphere}) with radius, $r$, is 
\begin{equation}
x^{2} + y^{2} + z^{2} = r^{2}.
\label{equ:p1} 
\end{equation}
Applying spherical coordinate transformation
\begin{equation}
  \begin{cases}
    x = r{}\sin{\theta}{}\cos{\varphi} \\
    y = r{}\sin{\theta}{}\sin{\varphi} \\
    z = r{}\cos{\theta}
     \end{cases}
  \label{equ:p2}    
\end{equation}
to the volume equation ($V = \frac{4}{3}{}\pi{}r^{3}$) of the sphere,
we can obtain the differential volume element
\begin{equation}
dV = r^{2}\sin{\theta}{}d{\theta}{}dr{}d{\varphi}.
\label{equ:p3} 
\end{equation}
$0{}\le{}\theta{}\le{}\pi$ and $0{}\le{}\varphi{}\le{}2{}\pi$
are shown in figure. 
The density profile for the sphere is
\begin{equation}
\rho = r^{-{\cal{}A}}.
\label{equ:p4} 
\end{equation}
Then the moment of inertia of the sphere about z axis can be calculated by
\begin{equation}
I = \int{}\int{}\int(x^{2} + y^{2}){\rho}\, dV \\
  = \int{}r^{4-{\cal{}A}}\,dr{}
      \int_{0}^{\pi}{}\sin^{3}{\theta}{}d{\theta}{}
      \int_{0}^{2 {}\pi}\,d{\varphi}.
\label{equ:p5} 
\end{equation}
\begin{equation}
\int_{0}^{\pi}{}\sin^{3}{\theta}{}d{\theta}= \frac{4}{3},
\label{equ:p6} 
\end{equation}
and
\begin{equation}
\int_{0}^{2{}\pi}\,d{\varphi} = 2 {}\pi
\label{equ:p7} 
\end{equation}
are easily getted.
Consequently,
\begin{equation}
I = \frac{8{}\pi}{3(5-{\cal{}A})}{}r^{5-{\cal{}A}} \\
  =\frac{2(3-{\cal{}A})}{3(5-{\cal{}A})}m{}r^{2},
\label{equ:p8} 
\end{equation}
where
\begin{equation}
m = \int{}\rho{}dV = \int{}r^{2-{\cal{}A}}\,dr{}
\int_{0}^{\pi}{}\sin{\theta}{}d{\theta}
 \int_{0}^{2 {}\pi}\,d{\varphi} \\
 = \frac{4{}\pi}{3-{\cal{}A}}{}r^{3-{\cal{}A}}.
\label{equ:p9}
\end{equation}

\begin{table*}
   \caption{Properties of the fitted clumps observed in 2MASS extinction map. \label{tab:clump}}
    \vspace*{0.5ex}
   \begin{tabular}{l c c  c l c l c l c l c c c c c }
   \hline
   \hline
     clumps & {} & R.A. , Decl. (J20000)  & {} & $M$  & {} & $R$  & {} & $R_{major}$ & 
     {} & $R_{minor}$ & {} & Angle & {} & $f_{obs}$ & $\gamma$ \\
     {}  & {} & (h:m:s, $^\circ$:$\arcmin$:$\arcsec$) & {} &  (M$_\sun$) & {} & (pc) & 
     {} & (pc) & {} & (pc) & {} & (deg E of N) & {} & {} & \\
     \hline
L1544a  & {}  & 05:04:24.21,+25:13:53.1 &  {}  &   40.8  & {}  &   0.14  & {}  &   0.20 & {}   &   0.09 & {} & -17  & {} &  0.47 & 1.29 \\
L1544b  & {}  & 05:03:46.96,+25:21:29.4 &  {}  &   39.0  & {}  &   0.11  & {}  &   0.14 & {}   &   0.09 & {} & 65  & {} &  0.65  & 1.19 \\
L1523   & {}  & 05:06:16.32,+31:43:58.6 &  {}  &   65.4  & {}  &   0.12  & {}  &   0.14 & {}   &   0.10 & {} & -102 & {} &  0.69 & 1.17 \\
B227a  & {}  & 06:07:27.06,+19:29:33.2 &  {}  &   50.7  & {}  &   0.23  & {}  &   0.25 & {}   &   0.21 & {} & -57  & {} &  0.81 & 1.10 \\
B227b  & {}  & 06:07:44.89,+19:28:14.7 &  {}  &    6.5  & {}  &   0.15  & {}  &   0.19 & {}   &   0.12 & {} & 109 & {} &  0.64  & 1.20 \\
L1574a  & {}  & 06:08:13.02,+18:31:49.0 &  {}  &   76.4  & {}  &   0.27  & {}  &   0.40 & {}   &   0.19 & {} &  -110 & {} & 0.47 & 1.30 \\
L1574b  & {}  & 06:08:02.69,+18:38:29.9 &  {}  &   54.7  & {}  &   0.19  & {}  &   0.25 & {}   &   0.15 & {} & -120 & {} &  0.61 & 1.21 \\
CB45a  & {}  & 06:08:44.25,+18:07:29.1 &  {}  &   88.4  & {}  &   0.24  & {}  &   0.29 & {}   &   0.20 & {} & 147 & {} &  0.69  & 1.17 \\
CB45b  & {}  & 06:08:58.87,+17:47:43.2 &  {}  &   34.6  & {}  &   0.18  & {}  &   0.21 & {}   &   0.15 & {} & 55  & {} &  0.73  & 1.14 \\
CB45c  & {}  & 06:08:42.05,+17:53:31.7 &  {}  &    4.1  & {}  &   0.12  & {}  &   0.16 & {}   &   0.09 & {} & 37  & {} &  0.57  & 1.23\\ 
L1257a  & {}  & 23:57:26.04,+59:40:14.7 &  {}  &    3.4  & {}  &   0.07  & {}  &   0.07 & {}   &   0.06 & {} & -2   & {} &  0.92 & 1.04 \\
L1257b  & {}  & 23:58:04.72,+59:36:48.5 &  {}  &    2.5  & {}  &   0.06  & {}  &   0.08 & {}   &   0.05 & {} & -9   & {} &  0.64 & 1.19 \\
\hline
\end{tabular}

\footnotesize {\rm\bf Notes.} Columns are (1) identified clumps, (2) coordinates of clumps center position, (3) mass of clumps, (4) clump radius, (5) clump semi-major axis, (6) clump semi-minor axis,(7) clump orientations, (8) clump axis ratio, (9) enhancement factor (see Equation~(\ref{prolate}) in Section~\ref{sec:ratio} and ~\citet{di2013}).
 \end{table*}

\begin{table*}
\centering
\caption{Results of spectral line fitting. \label{tab:width}} 
\begin{tabular}{lcccccccccc}
\hline
\hline 
 \multicolumn{6}{c}{$^{13}$CO}  & \multicolumn{5}{c}{C$^{18}$O} \\
 \cline{3-6} 
 \cline{8-11}
{} & {} & $T_{A}^{*}$ & $V_{LSR}$ & $\Delta{}V_{FWHM}$ & $\tau$ & {} & 
$T_{A}^{*}$ & $V_{LSR}$ & $\Delta{}V_{FWHM}$ & $\tau$\\
 clumps & {} & K & (km/s) & (km/s) &  {} & {} & K & (km/s) & (km/s)  & {} \\
 \hline
L1544a & {} & 1.34  $\pm$  0.07 & -0.87   $\pm$  0.02  & 0.83    $\pm$  0.04 &  0.32  & {}& 0.16   $\pm$    0.01 & -0.35   $\pm$  0.03 & 0.70   $\pm$  0.08  &  0.03 \\ 
L1544b & {} & 1.73  $\pm$  0.05 & -0.93   $\pm$  0.01  & 0.87    $\pm$  0.03 &  0.44  & {}& 0.13   $\pm$    0.01 & -0.31   $\pm$  0.04 & 0.77   $\pm$  0.09  &  0.03 \\ 
L1523  & {} & 2.40  $\pm$  0.03 & -0.44   $\pm$  0.01  & 1.48    $\pm$  0.01 &  0.67  & {}& 0.23   $\pm$    0.02 & -0.48   $\pm$  0.02 & 0.63   $\pm$  0.04  &  0.05 \\ 
B227a  & {} & 2.42  $\pm$  0.02 &  0.72   $\pm$  0.01  & 1.33    $\pm$  0.01 &  0.68  & {}& 0.24   $\pm$    0.02 & -0.25   $\pm$  0.01 & 0.72   $\pm$  0.03  &  0.05 \\ 
B227b  & {} & 1.59  $\pm$  0.02 &  0.35   $\pm$  0.01  & 1.58    $\pm$  0.01 &  0.39  & {} & 0.46   $\pm$   0.02 & -0.55   $\pm$  0.01 & 0.83   $\pm$  0.03  &  0.09 \\          
L1574a & {} & 3.30  $\pm$  0.04  & -0.43   $\pm$  0.01  & 1.09    $\pm$  0.01 &  1.12  & {}& 0.35   $\pm$  0.01 & -0.51   $\pm$  0.01 & 0.59   $\pm$  0.02  &  0.07 \\ 
L1574b & {} & 2.61  $\pm$  0.24 & -0.31   $\pm$  0.03  & 0.86    $\pm$  0.04 &  0.76  & {}& 0.09   $\pm$    0.01 & -0.60   $\pm$  0.07 & 0.90   $\pm$  0.19  &  0.02 \\ 
CB45a  & {} & 3.84  $\pm$  0.09 &  7.02   $\pm$  0.01  & 0.53    $\pm$  0.02 &  1.54  & {}& 0.56   $\pm$    0.02 & 7.01   $\pm$  0.01 & 0.30   $\pm$  0.01  &  0.11 \\ 
CB45b  & {} & 3.93  $\pm$  0.03 &  7.04   $\pm$  0.01  & 0.67    $\pm$  0.01 &  1.62  & {}& 0.45   $\pm$    0.02 & 6.92   $\pm$  0.01 & 0.37   $\pm$  0.01  &  0.09 \\ 
CB45c  & {} & 3.64  $\pm$  0.03 &  7.14   $\pm$  0.01  & 0.66    $\pm$  0.01 &  1.36  & {}& 0.46   $\pm$   0.02 & 7.11   $\pm$  0.01 & 0.45   $\pm$  0.01  &  0.09 \\ 
L1257a & {} & 2.74  $\pm$  0.01 & -0.69   $\pm$  0.01  & 1.47    $\pm$  0.01 &  0.82  & {}& 0.44   $\pm$    0.01 & -0.77   $\pm$  0.01 & 1.12   $\pm$  0.02  &  0.09 \\ 
L1257b & {} & 1.98  $\pm$  0.03 & -0.30   $\pm$  0.01  & 1.30    $\pm$  0.02 &  0.52  & {}& 0.27  $\pm$ 0.02 & -0.44 $\pm$  0.04 & 1.23   $\pm$  0.08  &  0.06 \\ 
 \hline
    \end{tabular}   
\end{table*}

\begin{table*}
\centering
\caption{Results of velocity gradient fitting. \label{tab:gradient1}} 
\begin{tabular}{lcccccccccccccccc}
\hline
\hline 
 & {} & {} & {} & \multicolumn{5}{c}{$^{13}$CO} & {} & {} & {} & \multicolumn{5}{c}{C$^{18}$O} \\ 
 \cline{3-9} 
 \cline{11-17}
{} & {} & N & {} & $\cal G$ & {} & {} & {} & $\theta_{\cal G}$ & {} & N & {} & $\cal G$ & {} & {} & {} & $\theta_{\cal G}$\\
 clumps & {} & {} & {} & (km/s/pc) & {} & {} & {} & (deg E of N) & {} & {} & {} & (km/s/pc) & {} & {} & {} & (deg E of N)\\
 \hline
L1544a & {} & 360 & {}  & 0.53    $\pm$  0.13 & {}& {} & {}  &  73   $\pm$    2   & {}& 84 & {}  & 1.42    $\pm$  0.47  & {} & {} & {}   &  79  $\pm$   9  \\
L1544b & {} & 364 & {}  & 0.54    $\pm$  0.13 & {}& {} & {}  &  77   $\pm$    8   & {}& 408 & {}  & 1.37    $\pm$  0.48  & {} & {} & {}   &  78  $\pm$   11 \\
L1523  & {} & 63 & {}  & 0.34    $\pm$  0.10 & {}& {} & {}  &  91   $\pm$    6   & {}& 58 & {}  & 0.50    $\pm$  0.16  & {} & {} & {}   & -89  $\pm$   4  \\
B227a  & {} & 176 & {}  & 0.41    $\pm$  0.11 & {}& {} & {}  &  10   $\pm$    8   & {}& 49 & {}  & 0.69    $\pm$  0.17  & {} & {} & {}   &  -3  $\pm$    7 \\
B227b  & {} & 126 & {}  & 1.15    $\pm$  0.34 & {}& {} & {}  &  12   $\pm$   5    & {}& 6 & {}  &           ---         & {} & {} & {}   &   ---  \\        
L1574a & {} & 498 & {}  & 0.37    $\pm$  0.11 & {}& {} & {}  & -157  $\pm$  41    & {}& 147 & {}  & 0.51    $\pm$  0.14  & {} & {} & {}   & -162 $\pm$   51 \\
L1574b & {} & 390 & {}  & 0.72    $\pm$  0.19 & {}& {} & {}  & -160  $\pm$  55    & {}& 57 & {}  & 0.87    $\pm$  0.21  & {} & {} & {}   & -167 $\pm$   21 \\
CB45a  & {} & 1216 & {}  & 0.44    $\pm$  0.14 & {}& {} & {}  &  165  $\pm$  49    & {}& 85 & {}  & 0.63    $\pm$  0.16  & {} & {} & {}   &  136 $\pm$   38 \\
CB45b  & {} & 316 & {}  & 1.46    $\pm$  0.11 & {}& {} & {}  &  158  $\pm$  20    & {}& 84 & {}  & 1.65    $\pm$  0.47  & {} & {} & {}   &  160 $\pm$   68 \\
CB45c  & {} & 520 & {}  & 0.66    $\pm$  0.10 & {}& {} & {}  &  172  $\pm$  61    & {}& 132 & {}  & 0.80    $\pm$  0.24  & {} & {} & {}   &  175 $\pm$   31 \\
L1257a & {} & 143 & {}  & 1.21    $\pm$  0.12 & {}& {} & {}  &  112  $\pm$  79    & {}& 74 & {}  & 1.45    $\pm$  0.43  & {} & {} & {}   &  147 $\pm$   23 \\
L1257b & {} & 142 & {}  & 1.31    $\pm$  0.12 & {}& {} & {}  &  178  $\pm$  35    & {}& 36 & {}  & 1.59    $\pm$  0.30  & {} & {} & {}   &  169 $\pm$   46 \\
\hline
\end{tabular} 

\footnotesize {\rm\bf Notes.} 
--- Clump B227b has fewer than nine chosen C$^{18}$O spectral lines and in consequence could not be reliably fitted. \\
\textbf{N is the number of the chosen $^{13}$CO and C$^{18}$O spectral lines.} 
\end{table*}

\begin{table*}
\centering
\caption{Calculated $J/M$ and $\beta$ based on the fitted velocity gradients. \label{tab:momentum}} 
\begin{tabular}{lccccccccccc}
\hline
\hline 
 & {} & {} & \multicolumn{4}{c}{$^{13}$CO} & {} & {} & \multicolumn{3}{c}{C$^{18}$O} \\ 
 \cline{4-7} 
 \cline{10-12}
{} & {} & {} & $J/M$ & {} & {} & $\beta$ & {} & {} & $J/M$ & {} & {} $\beta$ \\
 clumps & {} & {} & (pc km/s) & {} & {} & {} & {} & {}  & (pc km/s) & {} & {}  \\
 \hline
L1544a & {} & {} & 0.0099  $\pm$  0.0003  & {}& {}  &  0.0012   $\pm$   0.0008  & {} & {}& 0.0078  $\pm$  0.0004 & {} & {}  0.0086    $\pm$  0.001  \\   
L1544b & {} & {} & 0.0034 $\pm$  0.0004 & {}& {}  &  0.0039   $\pm$   0.0004  & {} & {}& 0.0025 $\pm$  0.0002 & {} & {}  0.0025    $\pm$  0.0006  \\ 
L1523  & {} & {} & 0.0060  $\pm$  0.0004 & {}& {}  &  0.0014   $\pm$   0.0002  & {} & {}& 0.0083  $\pm$  0.0006 & {} & {}  0.0026    $\pm$  0.0005  \\ 
B227a & {} & {} & 0.012  $\pm$  0.0002  & {}& {}  &  0.0016   $\pm$   0.0002  & {} & {}& 0.015  $\pm$  0.0003 & {} & {}  0.0044    $\pm$  0.0004  \\ 
B227b & {} & {} & 0.0072  $\pm$  0.002  & {}& {}  &  0.011    $\pm$   0.005   & {} & {}& ---    & {} & {}  ---  \\                                  
L1574a & {} & {} & 0.025  $\pm$  0.002  & {}& {}  &  0.0014   $\pm$   0.0002  & {} & {}& 0.021  $\pm$  0.003 & {} & {}  0.0026    $\pm$  0.0004  \\ 
L1574b & {} & {} & 0.0073  $\pm$  0.002  & {}& {}  &  0.0025   $\pm$   0.0003  & {} & {}& 0.0088  $\pm$  0.0003 & {} & {}  0.0037    $\pm$  0.0004  \\ 
CB45a & {} & {} & 0.020  $\pm$  0.002  & {}& {}  &  0.0012   $\pm$   0.0002  & {} & {}& 0.017  $\pm$  0.003 & {} & {}  0.0024    $\pm$  0.0003  \\ 
CB45b & {} & {} & 0.0097  $\pm$  0.001  & {}& {}  &  0.014    $\pm$   0.001   & {} & {}& 0.0080  $\pm$  0.0004 & {} & {}  0.018    $\pm$  0.003  \\   
CB45c & {} & {} & 0.0047  $\pm$  0.0004 & {}& {}  &  0.0071    $\pm$   0.004   & {} & {}& 0.0069  $\pm$  0.001 & {} & {}  0.010    $\pm$  0.002  \\   
L1257a & {} & {} & 0.0030 $\pm$  0.0001 & {}& {}  &  0.0046    $\pm$   0.0001  & {} & {}& 0.0041 $\pm$  0.0005 & {} & {} 0.0065     $\pm$  0.001  \\  
L1257b & {} & {} & 0.0022 $\pm$  0.0001 & {}& {}  &  0.069    $\pm$   0.001   & {} & {}& 0.0025 $\pm$  0.0004 & {} & {} 0.010     $\pm$  0.0007  \\ 
 \hline
    \end{tabular}   
\end{table*}    

\end{document}